\documentclass[%
reprint,
superscriptaddress,
showpacs,
preprintnumbers,
 bibnotes,
 amsmath,amssymb,
 aps,
prl
]{revtex4-1}

\usepackage{graphicx}
\usepackage{dcolumn}
\usepackage{bm}
\usepackage{color}

\begin{document}


\title{Preferred magnetic excitations in {Sr$_{1-x}$Na$_x$Fe$_2$As$_2$}}

\author{Jianqing~Guo}
\thanks{These authors contributed equally to this study.}
\affiliation{International Center for Quantum Materials, School of Physics, Peking University, Beijing 100871, China}
\author{Li~Yue}
\thanks{These authors contributed equally to this study.}
\affiliation{International Center for Quantum Materials, School of Physics, Peking University, Beijing 100871, China}
\author{Kazuki~Iida}
\affiliation{Neutron Science and Technology Center, Comprehensive Research Organization for Science and Society (CROSS), Tokai, Ibaraki 319-1106, Japan}
\author{Kazuya~Kamazawa}
\affiliation{Neutron Science and Technology Center, Comprehensive Research Organization for Science and Society (CROSS), Tokai, Ibaraki 319-1106, Japan}
\author{Lei~Chen}
\affiliation{International Center for Quantum Materials, School of Physics, Peking University, Beijing 100871, China}
\author{Tingting~Han}
\affiliation{International Center for Quantum Materials, School of Physics, Peking University, Beijing 100871, China}
\author{Yan~Zhang}
\affiliation{International Center for Quantum Materials, School of Physics, Peking University, Beijing 100871, China}
\affiliation{Collaborative Innovation Center of Quantum Matter, Beijing 100871, China}
\author{Yuan~Li}
\email[]{yuan.li@pku.edu.cn}
\affiliation{International Center for Quantum Materials, School of Physics, Peking University, Beijing 100871, China}
\affiliation{Collaborative Innovation Center of Quantum Matter, Beijing 100871, China}

\begin{abstract}
We have used inelastic neutron scattering to determine magnetic excitations in a single-crystal sample of {Sr$_{1-x}$Na$_x$Fe$_2$As$_2$}. The material's two magnetic phases, which differ in their orthorhombic and tetragonal lattice symmetries, share very similar strengths of magnetic interactions as seen from the high-energy excitations. At low energies, excitations polarized along the $c$ axis are suppressed in the tetragonal magnetic phase, in accordance with the associated reorientation of the ordered moments. Although excitations perpendicular to the $c$ axis are still prominent, only the weak $c$-axis response exhibits a spin resonant mode in the superconducting state. Our result suggests that $c$-axis polarized magnetic excitations are important for the formation of the superconductivity, and naturally explains why the critical temperature is suppressed in the tetragonal magnetic phase.

\end{abstract}


\maketitle

Magnetism is widely believed to be related to superconductivity in unconventional superconductors \cite{ScalapinoRMP2012}. In copper- and iron-based superconductors, magnetic correlations are also related to the formation of other electronic phases, in particular the spontaneous breaking of the lattice four-fold rotational symmetry \cite{HinkovScience2008,LuScience2014}, also known as the nematic order \cite{FradkinAnnuRev2010,FernandesNatPhys2014}. The multi-orbital nature of iron-based superconductors further enriches the manifestation of magnetism, since electrons' unquenched orbital angular momentum allows for a direct influence of spin-orbit coupling (SOC) on the magnetic ground and excited states. The energy scale of SOC in these materials is non-negligible \cite{BorisenkoNatPhys2015,WatsonPhysRevB2015,JohnsonPhysRevLett2015} compared to the typical phase transition temperatures. Indeed, inclusion of SOC into theoretical models has already allowed researchers to reproduce some of the most intricate electronic ground-state phase behaviors seen in experiments \cite{SchererCondmat2017,ChristensenCondmat2017A,ChristensenCondmat2017B}, and the influence of SOC on magnetic excitations has been observed with inelastic neutron scattering (INS) in both undoped \cite{QureshiPRB2012,WangPRX2013,MaPRX2017} and doped systems \cite{LuoPRL2013,ZhangPRB2013,QureshiPRB2014,SongPRB2016,HuPRB2017}. In spite of these efforts and the widely accepted notion that magnetic excitations may mediate Cooper pairing \cite{ScalapinoRMP2012,DaiRevModPhys2015}, how SOC might affect the formation of superconductivity has remained largely unexplored.

Hole-doped iron pnictides offer an opportunity to address this question in the so-called tetragonal magnetic phases \cite{AvciNatCommu2014,AllredPRB2015,TaddeiPRB2016,TaddeiPRB2017,MeierNPJQM2018}. These interesting phases are found over rather restrictive ranges of doping, inside the more common stripe-antiferromagnetic phase with a reentrant phase behavior -- the lattice symmetry is restored from orthorhombic ($C_2$) back to tetragonal ($C_4$). Although dependent on structural details \cite{MeierNPJQM2018}, in most cases the transition into the $C_4$-magnetic phase involves a reorientation of the ordered moments from the $ab$ plane to the $c$ axis \cite{AllredPRB2015,WasserPRB2015,MallettEPL2015,AllredNatPhys2016}. This indicates that magnetic anisotropy, one of the consequences of SOC, is important for selecting the ground state among possible contenders \cite{SchererCondmat2017,ChristensenCondmat2017A,ChristensenCondmat2017B}. Moreover, some of these $C_4$-magnetic phases compete strongly with superconductivity and suppress the critical temperature $T_\mathrm{c}$ \cite{BohmerNatCommun2015,WangPRB2016,TaddeiPRB2016}. Investigating how magnetic excitations change upon cooling into the $C_4$-magnetic and then the superconducting phases can thus be expected to reveal how SOC plays out in the microscopic mechanisms.

Motivated by the above considerations, here we use INS to determine spin excitations in hole-doped pnictide {Sr$_{1-x}$Na$_x$Fe$_2$As$_2$}. Over the doping range of $0.30 < x < 0.42$, this compound family possesses a relatively robust $C_4$-magnetic phase, accompanied by $c$-axis spin reorientation \cite{AllredNatPhys2016} and suppression of $T_\mathrm{c}$ \cite{TaddeiPRB2016}. Consistent with the notion that the $C_2$-to-$C_4$ magnetic phase transition is related to SOC \cite{SchererCondmat2017,ChristensenCondmat2017A,ChristensenCondmat2017B}, which is much lower in energy than the primary (isotropic) magnetic exchange interactions, we find that the main spectral change across the phase transition is at low energies, in that the $c$-axis-polarized response is strongly suppressed in the $C_4$-magnetic phase. Importantly, the suppression precludes the formation of a strong spin resonant mode \cite{ChristiansonNature2008,DaiRevModPhys2015,EschrigAdvPhys2006} in the superconducting state, because the resonant mode is found to be preferentially $c$-axis-polarized as well. This naturally explains why $T_\mathrm{c}$ is suppressed in the $C_4$-magnetic phase.

\begin{figure}
\includegraphics[width=3in]{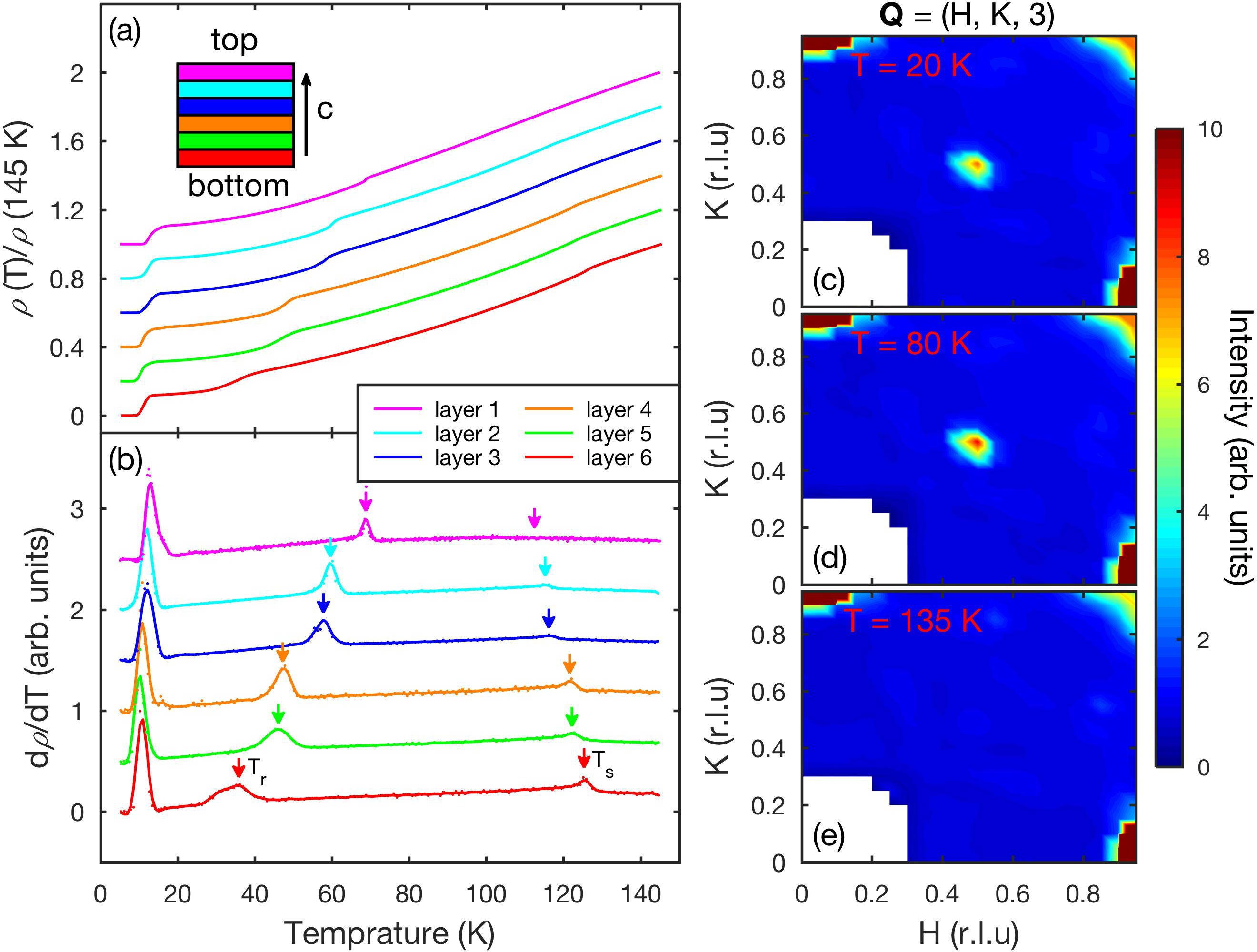}
\caption{\label{Fig1}
(a, b) Resistivity and the temperature derivatives measured on thin crystal flakes cleaved sequentially from a thick crystal (inset). Data are offset for clarity. Transition temperatures ($T_\mathrm{S}$, from paramagnetic to $C_2$-magnetic, and $T_\mathrm{r}$, from $C_2$-magnetic to $C_4$-magnetic) are indicated by arrows. (c-e) Neutron diffraction data acquired with $E_\mathrm{i} = $23 meV at different temperatures, after integrating over energy transfer 0-1.2 meV and momentum transfer along $\mathbf{c^*}$ 2.7-3.3 r.l.u.}
\label{Fig1}
\end{figure}

The single crystals of {Sr$_{1-x}$Na$_x$Fe$_2$As$_2$} used in our study were grown by a self-flux method \cite{TaddeiPRB2016}. As the INS experiment required a large sample mass, we took special care in checking the inhomogeneity of sodium concentration in thick crystals [Fig.~\ref{Fig1}(a,b)]. To ensure that the entire sample was in the desired electronic phases, crystal flakes were cleaved off both sides of each candidate crystal, characterized with resistivity measurements, and only crystals that showed clear signatures of both the $C_2$-to-$C_4$ magnetic phase transition and superconductivity on both sides were used for the INS experiment. A total of thirty such crystals were selected and co-aligned on aluminum plates using a hydrogen-free adhesive (see Fig.~S1 in \cite{SM}), amounting to a total mass of 4.5 grams and a mosaic spread of less than 4 degrees.  The INS measurements were performed at four temperatures: 135, 80, 20, and 6 K, at which the entire sample was in the paramagnetic, $C_2$-magnetic, $C_4$-magnetic, and superconducting phases (see Fig.~S1 in \cite{SM}), respectively.

Our INS experiment was performed on the 4SEASONS time-of-flight spectrometer at the MLF, J-PARC, Japan \cite{KajimotoJPSJ2011}. A chopper frequency of 250 Hz was used, and the spectrometer's multiple-$E_\mathrm{i}$ capability \cite{NakamuraJPSJ2009} allowed data in different energy ranges to be acquired simultaneously. During the measurements, the sample was rotated about the vertical axis over $\theta \in \left[-60,\,60\right]$ degrees in steps of $0.5^\circ$, where $\theta=0^\circ$ corresponds to $c$-axis orientation parallel to the incident beam. The acquired data were combined into a four-dimensional data set, which was analyzed with the Utsusemi \cite{InamuraJPSJ2013} and Horace \cite{EwingsNIMA2016} software packages. Figure~\ref{Fig1}(c-e) displays elastic scattering data taken in the $C_4$-magnetic, $C_2$-magnetic, and paramagnetic phases, respectively. Under the tetragonal notation which is used throughout this paper, a clear magnetic diffraction peak is seen in both magnetic phases at momentum transfer $\mathbf{Q} = (0.5, 0.5, 3.0)$ in reciprocal lattice units (r.l.u.). The slight decrease of the peak in the $C_4$-magnetic phase is caused by the spin reorientation \cite{AllredNatPhys2016} -- neutron scattering detects magnetic moments perpendicular to $\mathbf{Q}$, so that moments parallel to $c$ produce a smaller signal at this $\mathbf{Q}$ than in-plane moments.

\begin{figure*}
\includegraphics[width=4.375in]{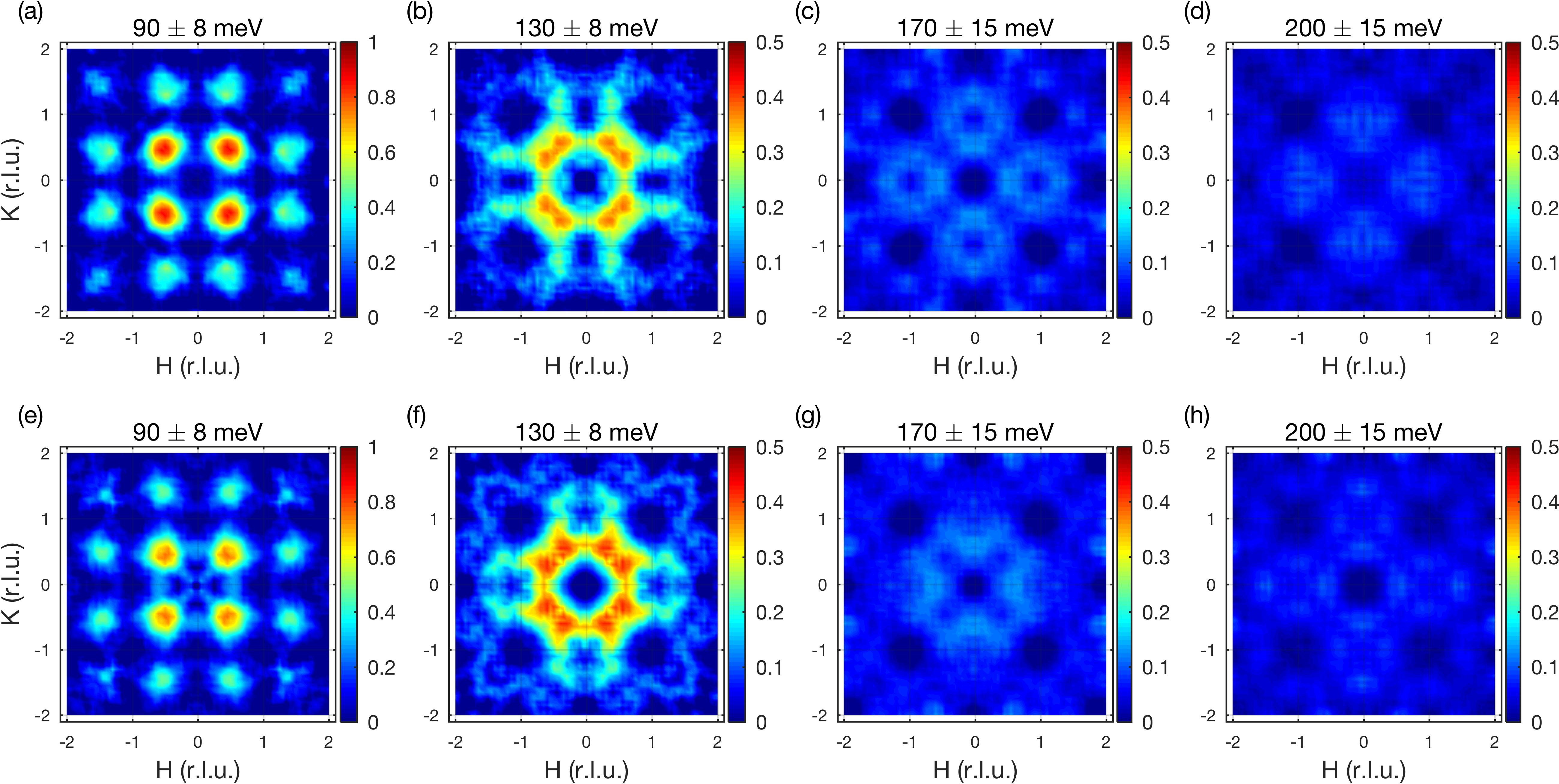}
\caption{\label{Fig2}
Constant-energy slices of magnetic excitations measured at 20 K (a-d) and 80 K (e-h), in the $C_2$- and $C_4$-magnetic phases, respectively. The measurements were performed with $E_\mathrm{i}=$ 320 meV, and data are plotted in the same arbitrary scattering cross-section units in all panels. The data have been integrated along the $c^*$ direction because of the quasi-two-dimensional nature of these high-energy excitations, fully symmetrized according to the $D_{4h}$ lattice symmetry, and smoothed for improved visualization. In the $C_2$ phase, the symmetrization is justified assuming an equal population of twin domains. Intensities integrated over $0.8<H<1.2$ and $0.8<K<1.2$ have been subtracted as background \cite{LiuNatPhys2012}. Data at other energies and temperatures are displayed in Fig.~S2 in \cite{SM}.}
\label{Fig2}
\end{figure*}

We first present our INS spectra measured at high energies. In the $C_2$-magnetic phase, the electronic structure is known to exhibit a polarization between the Fe $d_{xz}$ and $d_{yz}$ orbitals \cite{YiPNAS2011}. One may therefore expect that, in a local-moment picture, the magnetic exchange interactions along the two in-plane Fe-Fe directions are different \cite{ZhaoNatPhys2009}, a situation that could fundamentally change upon entering the $C_4$-magnetic phase. Our INS data (Fig.~\ref{Fig2}) show that this is \textit{not} the case. The $C_2$- and the $C_4$-magnetic phases look very similar over a broad energy range, with the excitation band top being at the magnetic Brillouin zone corner, such as $(H, K) = (1, 0)$ in Fig.~\ref{Fig2}(d) and (h). The subtle differences towards the highest energy are statistically insignificant (see Fig.~S3 in \cite{SM}). Overall, the spectral shape is consistent with that in the undoped system \cite{ZhaoNatPhys2009} which has only the $C_2$-magnetic phase. Similar excitation spectra, with a bandwidth of about 200 meV, have recently been reported for hole-doped Ba$_{0.75}$K$_{0.25}$Fe$_2$As$_2$ \cite{MuraiPRB2018}.

Some cautions are noteworthy when considering the similarity in Fig.~\ref{Fig2} between the two phases: (1) the magnetic ground states are different, (2) our sample must be ``twinned'' in the $C_2$-magnetic phase, (3) the excitations may be best described from neither a purely local-moment nor a purely itinerant point of view \cite{WangPRX2013,ZhaoNatPhys2009,EwingsPRB2011,MuraiPRB2018,LuCondmat2018}, and (4) our INS spectra are ``orbitally summed'', whereas ``orbitally resolved'' measurements might reveal additional contrast between the two phases \cite{SchererPRB2016}. Nevertheless, our result suggests that the $C_2$-to-$C_4$ magnetic phase transition is not directly related to changes in the primary magnetic interactions (local-moment picture), or in the electronic structure far away from the Fermi level (itinerant picture). In fact, the high-energy spectrum remains roughly the same all the time, even in the paramagnetic phase (see Fig.~S2 in \cite{SM}).

\begin{figure}
\includegraphics[width=3in]{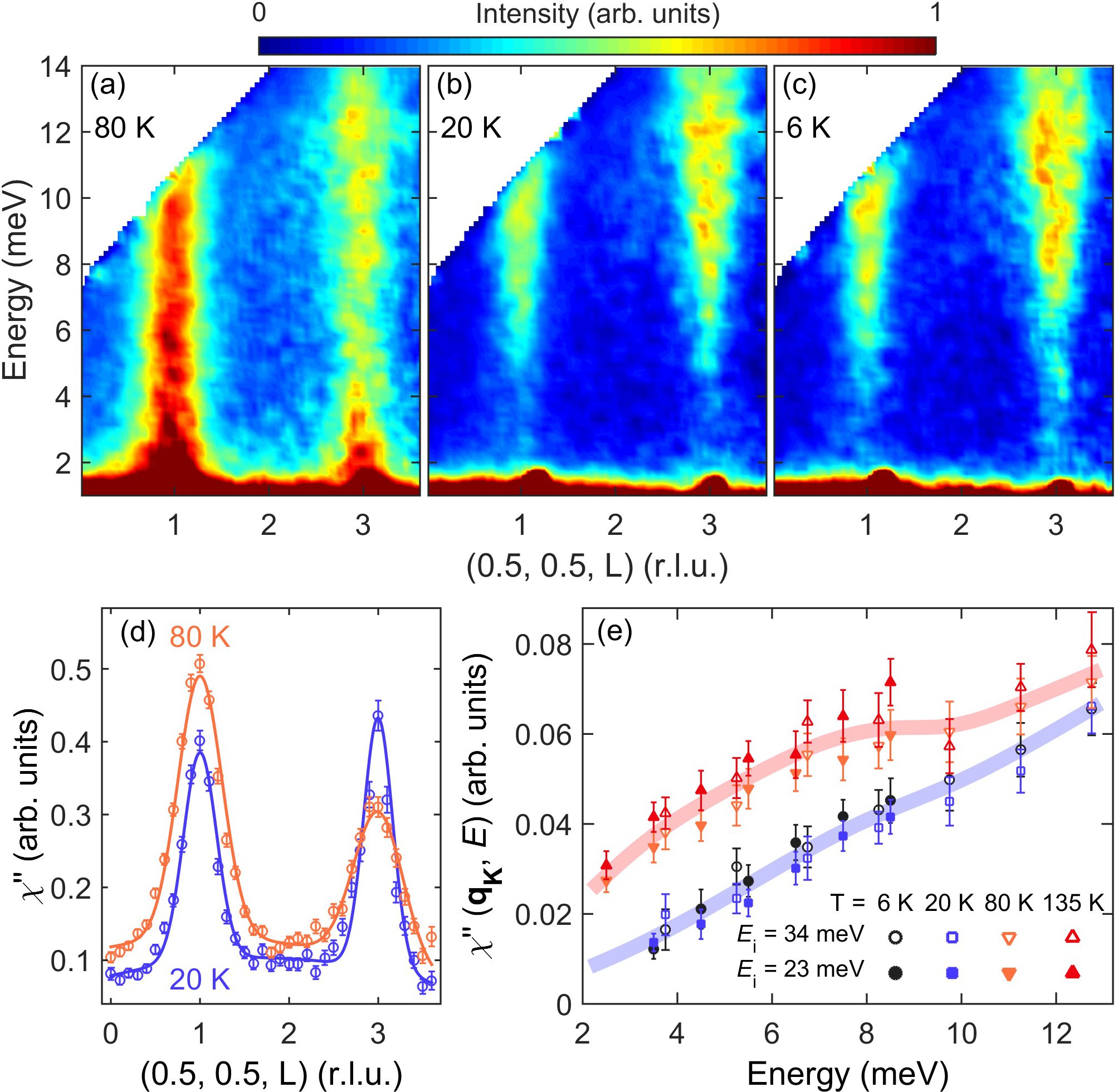}
\caption{\label{Fig3}
(a-c) INS signals integrated over $H,\,K \in [0.45, 0.55]$ as a function of $L$ and energy, measured with incident neutron energy $E_\mathrm{i} = 23$ meV at different temperatures. (d) Imaginary part of dynamic susceptibility (see text) integrated over 4-8 meV based on the data in (a) and (b). (e) Momentum-integrated susceptibility near (0.5, 0.5, 1) (see Fig.~S3 in \cite{SM} for more detail) measured at different temperatures.
}
\label{Fig3}
\end{figure}

In contrast, Fig.~\ref{Fig3}(a-b) shows a clear change upon entering the $C_4$-magnetic phase at low energies. The change further manifests itself differently at physically equivalent momentum transfers along $\mathbf{c^*}$, $L = 1$ and 3, which correspond to different angles between $\mathbf{Q}$ and $\mathbf{c^*}$. Figure \ref{Fig3}(d) displays the imaginary part of the dynamic spin susceptibility (proportional to the INS intensity divided by the Bose factor), integrated over 4-8 meV and a momentum range near the in-plane magnetic ordering wave vector. This quantity, once integrated near the integer $L$ values, decreases by about a factor of two upon cooling from 80 K to 20 K at $L=1$, but remains approximately unchanged at $L=3$ (apart from a smaller width in $L$ at 20 K). Since INS detects magnetic excitations polarized perpendicular to $\mathbf{Q}$, the above difference indicates that the suppression at $L=1$ results from a decrease in the $c$-axis polarized excitations in the $C_4$-magnetic phase. At $L=3$, the INS signal contains more in-plane than $c$-axis response, the former of which is expected to increase in the $C_4$-magnetic phase (see below), so the total signal exhibits no significant decrease or even a slight increase, depending on the energy. Taking into account all data for $L=1$ measured at four different temperatures [Fig.~\ref{Fig3}(e)], the susceptibility suppression is found to occur mainly between 80 and 20 K, and mainly below about 10 meV. This energy is somewhat smaller than a characteristic energy recently seen by electronic Raman scattering \cite{YuePRB2017} in the $C_4$-magnetic phase, but the discrepancy is not totally unexpected because Raman and INS probe different responses (charge and spin, respectively) of the electrons.

In a local-moment picture, low-energy spin excitations from a magnetically ordered state are dominated by spin waves, which are transverse to the direction of the ordered moments. Longitudinal excitations are forbidden when the system is far from any critical points. Since the ordered moments are along the $a$ and the $c$ axis in the $C_2$- and the $C_4$-magnetic phases \cite{AllredNatPhys2016}, respectively, the in-plane excitations are expected to be enhanced at the cost of suppression of the $c$-axis excitations in the $C_4$-magnetic phase, consistent with our result. The enhancement of in-plane response in the $C_4$-magnetic phase is also consistent with large nematic susceptibility observed in this phase \cite{WangPRB2018}. Meanwhile, it is known that the low-energy spin excitations have an itinerant character \cite{WangPRX2013}, which is further confirmed by the fact that superconductivity fundamentally modifies the excitation spectrum by forming the so-called spin resonant mode(s) below $T_\mathrm{c}$ \cite{ChristiansonNature2008,DaiRevModPhys2015,EschrigAdvPhys2006}. Our result therefore suggests that both the local and the itinerant pictures are needed in order to fully account for the experimental data.

\begin{figure}
\includegraphics[width=3in]{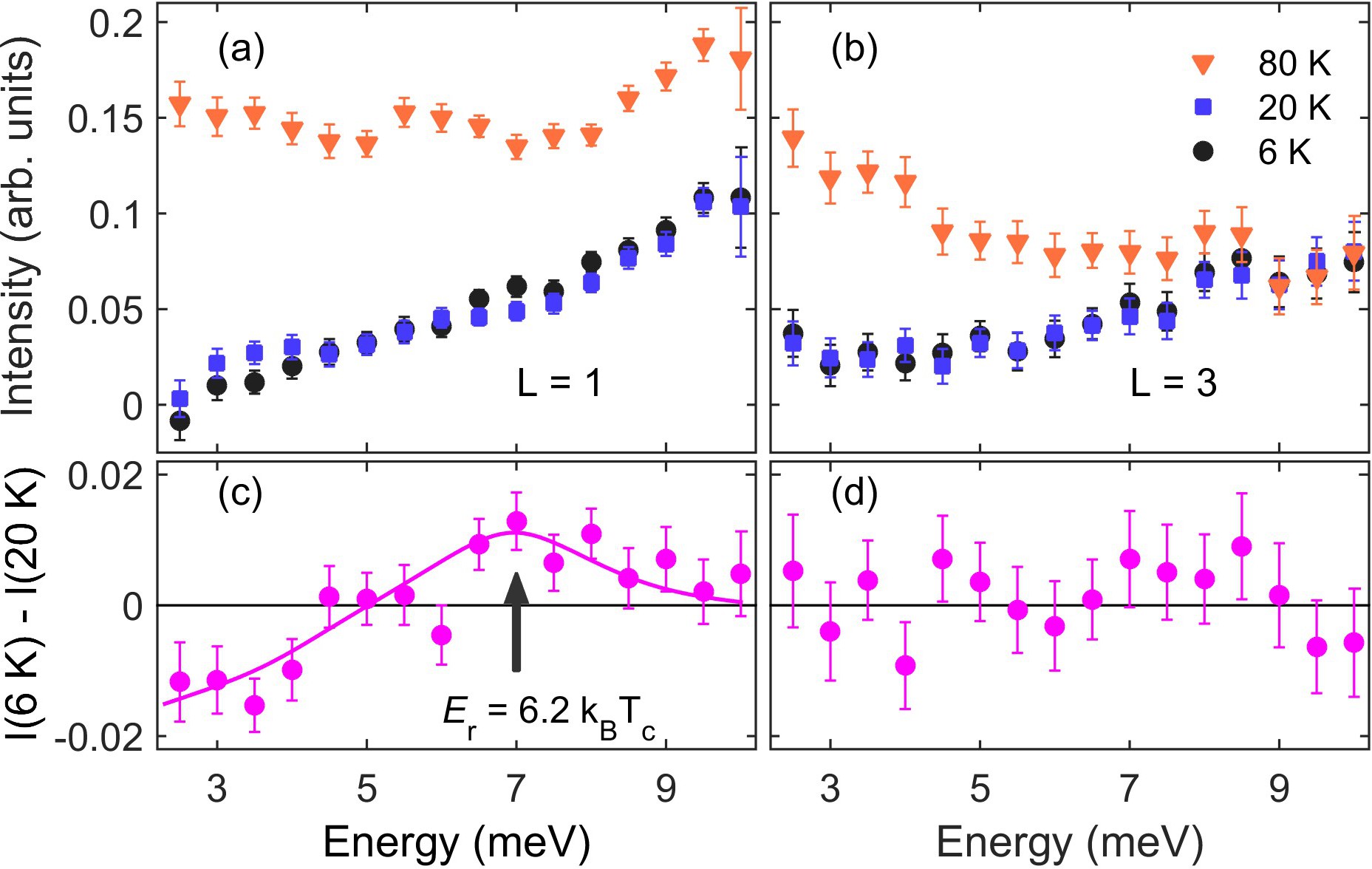}
\caption{\label{Fig4}
(a, b) Constant-$\mathbf{Q}$ cuts measured with $E_\mathrm{i}=23$ meV at $(0.5, 0.5, 1)$ and $(0.5, 0.5, 3)$. The averaged intensity at $(H, K)=(0.3, 0.3)$ and $(0.7, 0.7)$ has been subtracted as background. The same background measured at 20 K is subtracted from the 6 K and the 20 K data, whereas the 80 K data are compared against their own background. (c, d) Intensity difference between 6 K and 20 K at $L=1$ and 3.
}
\label{Fig4}
\end{figure}

A surprise in this regard is the lack of a prominent spin resonant mode in the data shown in Fig.~\ref{Fig3}(b-c). Resonant modes are observed in unconventional superconductors in which the Cooper pairing mechanism might be related to magnetic interactions \cite{EschrigAdvPhys2006,YuNatPhys2009,InosovPRB2011}. The absence of a prominent spin resonant mode in our sample is probably related \cite{WangNatCommun2013} to the fact that the superconducting condensation energy, as can be seen by the specific heat anomaly at $T_\mathrm{c}$, is suppressed over the doping range where the $C_4$-magnetic phase is present \cite{BohmerNatCommun2015,WangPRB2016}. We have therefore deliberately searched for the resonant mode, by closely comparing the intensities recorded below and above $T_\mathrm{c}$ at the magnetic wave vector, as shown in Fig.~\ref{Fig4}, for both $L=1$ and 3. Indeed, a weak enhancement is found at $L=1$ at about 7 meV $=6.2 k_\mathrm{B} T_\mathrm{c}$ (the average $T_\mathrm{c}$ of our sample is about 13 K), which is close to the expected energy seen in many iron-based superconductors \cite{InosovPRB2011}. Importantly, this intensity enhancement below $T_\mathrm{c}$, along with the intensity suppression at lower energies, is not observed at $L=3$, despite the fact that there are actually more low-energy magnetic signals there than at $L=1$. In other words, although plenty of in-plane polarized excitations are still present in the $C_4$-magnetic phase, they do \textit{not} participate in the formation of the spin resonant mode. This is a central aspect of our finding.

Assuming that magnetic excitations might serve as ``glue'' bosons for Cooper pairing, our result reveals that the $c$-axis polarized excitations might be the most relevant glue. This is because the spin resonant mode can be understood as a feedback effect of Cooper pairing on the spin excitations \cite{EschrigAdvPhys2006}, hence the most affected excitations are likely also those that interact most strongly with the fermionic quasiparticles. In our system, the $c$-axis polarized resonant mode has to be weak, because the $C_4$-magnetic phase suppresses $c$-axis polarized excitations as longitudinal excitations already in the non-superconducting state, leaving little spectral weight for the formation of the resonant mode below $T_\mathrm{c}$. The above line of thinking naturally explains why Cooper pairing is weakened in the presence of the $C_4$-magnetic phase. In conjunction with the fact that the $C_4$-magnetic phase possesses a non-uniform distribution of magnetic moment density on Fe seen by local probes \cite{AllredNatPhys2016}, our result points towards an important interplay between the itinerant charge carriers and the local spins on Fe, namely, a realization of ``Hund's metals'' \cite{YinNatMater2011}.

To further rationalize the experimental findings, we believe that the quasiparticle wavefunctions near the Fermi level, under the influence of SOC \cite{BorisenkoNatPhys2015,WatsonPhysRevB2015,JohnsonPhysRevLett2015}, introduce a nontrivial structure into the scattering matrix elements (concerning the spin polarizations) between the quasiparticles and the spin excitations \cite{MaPRX2017}. When the matrix elements are most compatible with the polarization of the excitations arising from the local moments, the scattering is strong and good for mediating Cooper pairing. The $C_4$-magnetic phase here is against this requirement. In contrast, the so-called hedgehog spin-vortex crystal phase, which has been observed in the ``1144'' pnictides \cite{MeierNPJQM2018}, is \textit{not} against this requirement because the ordered moments are in-plane, and indeed its presence does not seem to suppress superconductivity \cite{MeierNPJQM2018,DingPRB2017}. Last but not least, the spin resonant mode in many iron-based superconductors is preferentially polarized along the $c$ axis \cite{LuoPRL2013,MaPRX2017,SongPRB2016,ZhangPRB2013}, and the anisotropy only gradually disappears towards high doping \cite{QureshiPRB2014,LiuPRB2012}, where the increased kinetic energy reduces the SOC effects near the Fermi level, consistent with the above picture. Since our experiment shows that such spin anisotropy of the resonant mode persists even when the most nearby (in fact, coexisting) magnetic order is against it, it cannot be attributed to the stripe-antiferromagnetic state in the parent compound as was previously suggested \cite{LiuPRB2012,ZhangPRB2013,HuPRB2017}.

To summarize, we observe no significant difference in the primary magnetic interactions in the $C_2$- and $C_4$-magnetic phases. In spite of a clear suppression of $c$-axis polarized excitations in the $C_4$-magnetic phase, the resonant mode is still observed mostly in the $c$-axis response. Assuming that the anisotropy of the resonant mode is caused by matrix-element effects on itinerant electrons due to SOC, we attribute the suppression of $T_\mathrm{c}$ in the $C_4$-magnetic phase to an incompatible combination of preferred spin excitations from the itinerant and the local points of view. In Hund's metals with SOC, better superconductors may hence be obtained if such compatibility requirements are better satisfied.

\begin{acknowledgments}

We wish to thank F. Wang, D.-H. Lee, P. Dai, B. M. Andersen, F. Hardy, C. Meingast, and J. Schmalian for stimulating discussions. The work at Peking University is supported by the NSF of China (grant No. 11522429) and the NBRP of China (grant Nos. 2015CB921302 and 2018YFA0305602).

\end{acknowledgments}

\bibliography{Reference}

\end{document}